\documentclass[runningheads]{svmult}

\usepackage{makeidx}   
\usepackage{graphicx}  
\usepackage{subeqnar}  
\usepackage{multicol}  
\usepackage{physprbb}  
\makeindex             

\begin{document}
\title*{Spline histogram method for reconstruction of 
probability density functions of clusters of galaxies}
\toctitle{Spline histogram method for reconstruction of
probability density functions of clusters of galaxies}
%
%
\titlerunning{Spline histogram method}
%
\author{Dmitrijs Docenko\inst{1}
\and K\={a}rlis B\={e}rzi\c{n}\v{s} \inst{2}
}
\authorrunning{Dmitrijs Docenko and Karlis Berzins}
%
%
\institute{Institute of Astronomy, University of Latvia,
   Raina blvd 19, Riga LV-1586, Latvia; e-mail: dima@latnet.lv
\and Ventspils International Radio Astronomy Center,
   Akademijas laukums 1-1503, Riga LV-1050, Latvia; e-mail: kberzins@latnet.lv}

\maketitle              

\begin{abstract}
We describe the spline histogram algorithm which is useful for visualization 
of the probability density function setting up a statistical hypothesis for a 
test. The spline histogram is constructed from discrete data measurements 
using tensioned cubic spline interpolation of the cumulative distribution
function which is then differentiated and smoothed using the Savitzky-Golay
filter. The optimal width of the filter is determined by minimization of
the Integrated Square Error function.

The current distribution of the TCSplin algorithm written in f77 with IDL
and Gnuplot visualization scripts is available from
www.virac.lv/en/soft.html.
\end{abstract}

\section{Introduction}

Whenever one makes a physical measurement one obtains a discrete result,
starting from spatial measurements and ending with classification of some
set of objects by some quantity. Particular measured value follows from
the statistical properties of the system strictly following the probability
distribution function, hereafter PDF. The PDF, in its turn, is determined by
the physical properties of the system. If a measured data set is statistically
complete then its PDF contains information about the system's physical
properties. The PDF shows a character of unimodal or multimodal systems. It
is natural to assume that the PDF of unimodal physical systems contain
only one global maximum and several maxima indicate the multimodality of a
data set. Therefore the shape of the PDF allows one to classify the
measured data points, e.g.\ to find structure in case of positional
measurements.

In statistics it is widely accepted to use histograms as the PDF
approximations. It is also well known that ordinary histograms being
dependent on two free parameters (bin size and its location) give very
subjective results. Many methods have been developed trying to solve this 
problem \cite{KBMSc}. However, most of them are still dependent on some 
parameters in a non-objective manner.

Generally, the probability density estimation methods can be divided into
two main groups: parametric and non-parametric. The first ones assume some
definite type of the PDF function (e.g. Gaussian or their superposition) and
try to find the best-fit parameters for it. A good such example is the KMM
mixture modelling algorithm \cite{Ashman}. A main disadvantage of these methods
is that not all data sets can be well fitted with any chosen function.
Rather often the real PDF of physical system has significant difference
from a chosen best-fit function, and in many cases it is not known 
\emph{a priori} what function it should be at all.

Non-parametric methods try to construct PDF estimates as compromise of two
opposite demands. First, the estimate should be as close as possible to
the measured PDF. Second, statistical noise due to a finite volume of the
selection should be filtered out. There are several ways how to do it.

It is possible to minimize a functional that is a sum of two terms -- the 
statistical noise and the one increasing with a difference between data 
points and the PDF estimate (Vondrak's method) \cite{Vondrak}. 
Unfortunately, there is still one
free coefficient responsible for the smoothing degree. This coefficient
is not determined in any automated way and usually is found from
good-looking conditions. Another method is to convolve the initial guess of
the PDF with some kernel (kernel methods) for data smoothing. Also in
this case there remains a free coefficient -- the kernel width, that is 
responsible for smoothing of the function in an ``optimal way'', besides 
the result is weakly dependent on the chosen kernel shape 
\cite{KBMSc}, \cite{Vio}.

There is, however, a method that allows one to choose an optimal smoothing 
width: the PDF should not be over-smoothed and lose its true features, and 
the noise level should be diminished as far as possible on the other hand. 
This method is described e.g.\ in \cite{Vio}. Its main idea is to
define an Integrated Square Error ($ISE$) function that shows the
difference between the real PDF and its estimate, and then to minimize it.
The $ISE$ function method is implemented for kernel methods in e.g.\ 
\cite{Pisani1}, and results are encouraging. However, the $ISE$ function itself
is often rather noisy.

In this paper we propose another approach to estimate PDF of a given
one-dimensional data set in automatic and optimal way. This is the spline
histogram method. We have found that the tensioned cubic splines are
suitable for this task and the corresponding algorithm has been called
TCSplin. The current version of the TCSplin code is freely available in the
internet, it is also included in the CD-ROM.

For demonstration purposes in this article we will use the spline histogram
algorithm to find a ``well determined'' redshift structure of galaxies
within clusters Abell 2256 and Abell 3626.

This paper has the following structure. The spline histogram algorithm will
be discussed in section~\ref{sha}. Bootstrapping simulations, discussed in 
section~\ref{sda}, help to evaluate errors of the PDF estimates. In
section~\ref{clu} the spline histogram application to data sets of clusters 
of galaxies A2256 and A3526 will be shown as examples. 
Finally, some concluding remarks will be given in section~\ref{beigas}.

\section{The spline histogram algorithm}
\label{sha}

The spline histogram method is a non-parametric approach for reconstruction of
probability density function underlying statistical selection. It was
first discussed in \cite{KB0} as one of possible methods to detect
substructure in clusters of galaxies. Recently it was further developed in 
\cite{DDMSc} and these results are summarized in this paper.

From spectroscopic observations we obtain redshifts of galaxies in 
clusters. Let us denote redshift of the $i$th galaxy by $z_i$ and order them
ascendentally ($z_i \leq z_{i+1}$). Next step is to construct a step-like
cumulative distribution function (CDF) obtained purely from observational
data: $F_{obs}(cz)=N(z_{j}<z)/N_{gal}$, where $N(z_{j}<z)$ is a number of
galaxies with redshift smaller than $z$, and $N_{gal}$ is a total number of
detected galaxies, $c$ is the speed of light. At this stage we
assume that the data set is statistically complete being representative of
the physical situation. The PDF $f(x)$ by definition is the derivative of $%
F_{obs}(cz)$ in respect to $cz$. If CDF is constructed as shown
before then $f(x)$ is a sum of Dirac $\delta $-functions.

In the spline histogram method the points $z_{i}$ with ordinates $%
F_{obs}(cz_{i})$ are consequently connected by non-decreasing smooth
analytical spline $S(cz)$. After construction of $S(cz)$ the latter is
analytically differentiated leading to the PDF estimate $\hat{f}(cz)$. This
procedure guarantees that the obtained continuous PDF is in agreement with
the discrete distribution of the data points. The PDF contains all initially
observed information about the cluster and it has a lot of
noise as a consequence. To diminish the noise, $\hat{f}(cz)$ has to be smoothed.

Trying to utilize usual cubic splines for interpolation of the CDF, one
encounters the problem that they will generally have negative derivative
intervals if both first and second derivatives in the data points are put
to be equal. Although there is an infinite amount of possibilities how to
construct a smooth continuous spline that has non-smooth first derivative
at data points.

We have found that tensioned cubic splines (hereafter TCS) nicely fit all
spline histogram needs. They are defined such that the cubic polynomial
spline length between two data points is minimal, and only the interpolating
function and its first derivative are continuous in data points. Also in
this case sometimes a derivative of the TCS is negative. Then in order to
exclude a non-physical decreasing of the CDF estimate, we use non-tensioned
splines increasing accordingly the spline length within these regions.

To reduce the statistical noise, the algorithm has been symmetrized. For the
same purpose there was added a possibility to unite close points in the data
set, that would otherwise give unphysical high PDF peaks. Nevertheless the
resulting PDF is 
noisy and due to this the next step of the algorithm is a smoothing
procedure.

In our case the noise is seen as narrow high peaks in the PDF arising from
high CDF derivatives between close data points. 
The Savitzky-Golay filters \cite{NR} have been chosen for the smoothing
remembering that PDF construction without any \emph{a priori} knowledge 
about the system character was one of the main reasons for developing 
the spline histograms.
These filters locally conserve first moments of the smoothed function. The
remaining problem is to define the optimal width of the filter such
that it reduces the noise but not over-smoothes the real PDF features.

This is done using the Integrated Square Error ($ISE$) function \cite{Vio}%
: 
\begin{equation}
ISE(\hat{f}(cz))=\int\limits_{cz_{min}}^{cz_{max}}\left( \hat{f}%
(cz)-f(cz)\right) ^{2}d(cz),
\end{equation}%
where $f(cz)$ is a true PDF underlying the observed selection, and $\hat{f}%
(cz)$ is a PDF estimated from the observed data, i.e.\ the smoothed spline
histogram in our case. As we are using digital filters to smooth the data,
this should be rewritten for case of discrete points. It follows from the
theory \cite{Silverman}, \cite{DDMSc}, \cite{Vio} that quantity $P(h)$ will
have minimum for the same smoothing width $h$ as $ISE(\hat{f}(x))$: 
\begin{equation}
P(h)=\sum\limits_{i=1}^{N}\left( \hat{f}(cz_{i})^{2}-2\hat{f}%
(cz_{i})+2C_{0}^{(h)}\right) ,
\end{equation}%
where $C_{0}^{(h)}$ are the smoothing filter zeroth coefficients, and it was
taken into account that $\sum_{i=1}^{N}\hat{f}(cz_{i})=N$. In contrary to
the equation defining the $ISE$ function, $P(h)$ can be easily calculated
from the data.

The filter width that gives the minimal $P(h)$ value is the optimal one
because the corresponding deviation between the true and estimated PDFs is
also minimized. As a result the spline histogram is obtained but it says
nothing about the remaining statistical noise level in it. To find it out we
use a bootstrapping technique described in the next section.

\section{Simulated data analysis}

\label{sda}

\begin{figure}[t]
\par
\begin{center}
\includegraphics[width=1\textwidth]{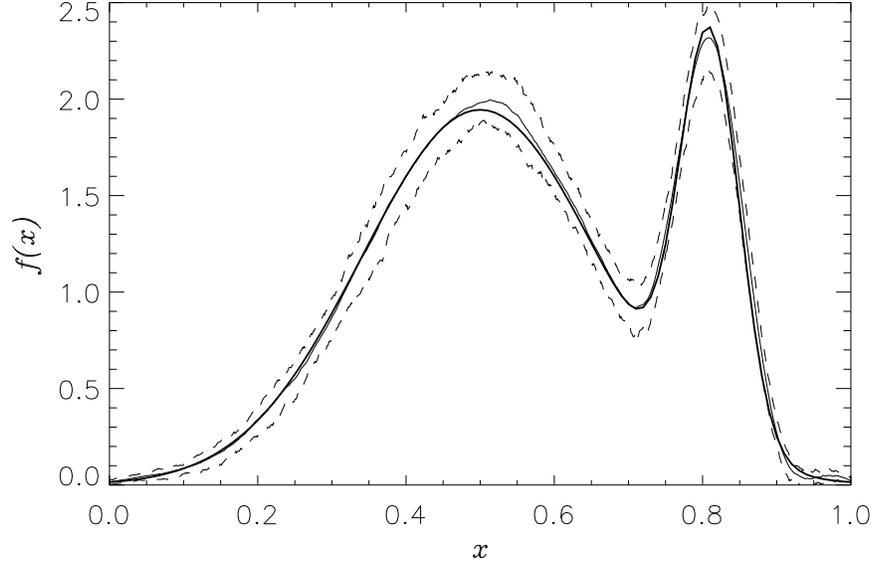}
\end{center}
\caption{Result of the simulated distribution analysis. Original PDF is
shown by the thick solid line, the thin solid line represents the average
value of 100 smoothed spline histograms using 500 point selection each, and
the lower and upper dashed lines are the first and third quartiles,
respectively. }
\label{sim500}
\end{figure}

Simulated data are produced and analysed as follows. Using the obtained
spline histogram as a true PDF, we generate the same amount $N$ of random
numbers. Then from this selection we compute another smoothed spline
histogram. Repeating this sufficient number of times (say 100), one can
calculate the average of the simulated spline histograms and its scattering.
It is useful to characterize the scattering by the distribution quartiles.
The upper quartile shows that the estimated PDF has 75\% probability to be
below it. For the lower quartile, accordingly, this probability is 25\% (see
Fig.~\ref{sim500}).

\begin{table}[hb]
\caption{Moments of the initial distribution from simulations of the 
500 point selection}
\label{sim}
\begin{center}
\renewcommand{\arraystretch}{1} \setlength\tabcolsep{5pt} 
\begin{tabular}{@{}lcccc}
\hline
\noalign{\smallskip} \textbf{Gaussian distribution} & Average & St.Dev. & 
Asymmetry & Excess \\ 
General distribution & 0.500 & 0.089 & \phantom{-}0.000 & -0.006 \\ 
Average from simulations & 0.499 & 0.093 & -0.486 & \phantom{-}2.287 \\ 
St.Dev. from simulations & 0.004 & 0.003 & \phantom{-}0.109 & \phantom{-}%
0.355 \\ \hline
\textbf{2 equal dispersion Gaussians} & Average & St.Dev. & Asymmetry & 
Excess \\ 
General distribution & 0.475 & 0.190 & \phantom{-}0.190 & -0.718 \\ 
Average from simulations & 0.474 & 0.192 & \phantom{-}0.153 & -0.676 \\ 
St.Dev. from simulations & 0.009 & 0.005 & \phantom{-}0.068 & \phantom{-}%
0.104 \\ \hline
\textbf{2 different dispersion Gaussians} & Average & St.Dev. & Asymmetry & 
Excess \\ 
General distribution & 0.565 & 0.190 & -0.112 & -0.836 \\ 
Average from simulations & 0.564 & 0.192 & -0.165 & -0.704 \\ 
St.Dev. from simulations & 0.009 & 0.004 & \phantom{-}0.064 & \phantom{-}%
0.105 \\ \hline
\end{tabular}%
\end{center}
\end{table}

To estimate the quality of the approximation, the first moments of several
simulated distributions were computed and compared with the original values
(see Table~\ref{sim}). It can be seen that the average values are the same
within statistical error bars ($1\sigma$), whereas the standard deviations
are about 10\% larger than the original values because of the smoothing
effect. For Gaussian distributions the asymmetry and excess are
significantly different from zero, although in non-Gaussian cases they are
rather close to original values.

Dependence of the smoothing size on the selection volume was also analysed.
From theoretical considerations \cite{Silverman}, \cite{Vio}, \cite{Pisani1},
we know that the optimal smoothing size depends on the selection volume $N$
in the following way: $h_{opt}\propto N^{-1/5}$. Analysing different volume
random number selections for the same initial distribution, we have
empirically found that for our algorithm $h_{opt}\propto N^{-0.195}$, that
shows an excellent agreement with the theoretical prediction.

\section{Galaxy cluster data analysis}

\label{clu}

\begin{figure}[t]
\par
\begin{center}
\includegraphics[width=1\textwidth]{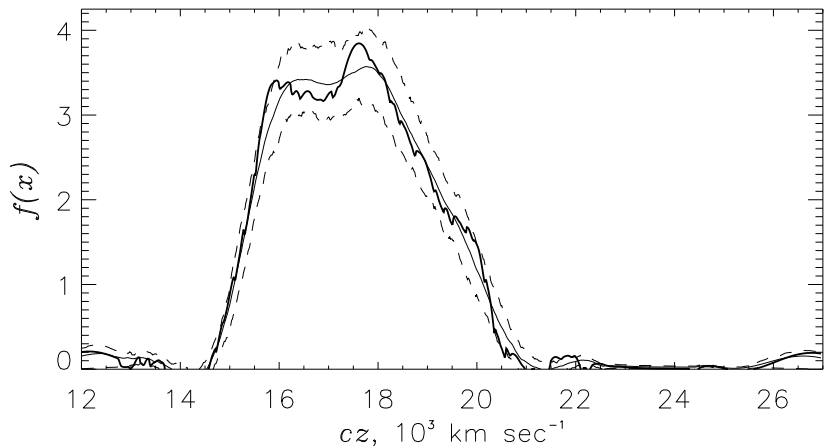}
\end{center}
\caption{The spline histogram of A2256. The meaning of lines is the same as
in Fig.~\protect\ref{sim500}}
\label{a2256}
\end{figure}

As example we show the implementation of the algorithm on two clusters of
galaxies -- Abell 2256 and 3526.

Abell 2256 is a rich regular cluster at $z\approx 0.06$ ($\alpha \approx
17^{h}03.7^{m}$, $\delta \approx +78^{\circ }43^{\prime }$, equinox 2000.0 
\cite{Abell}). It has similar 
properties to the Coma cluster (similar X-ray luminosities, both have
optical and X-ray substructure and a radio halo), but is situated
approximately 2.5 times farther.

A2256 has been previously studied in x-rays, optical and radio by several
authors, e.g. \cite{Fabricant}, \cite{Briel}, \cite{Clarke}. It is accepted
and understood, that Abell 2256, being one of the best studied
clusters of galaxies, exhibit complex inner structure.

\begin{figure}[t]
\par
\begin{center}
\includegraphics[width=1\textwidth]{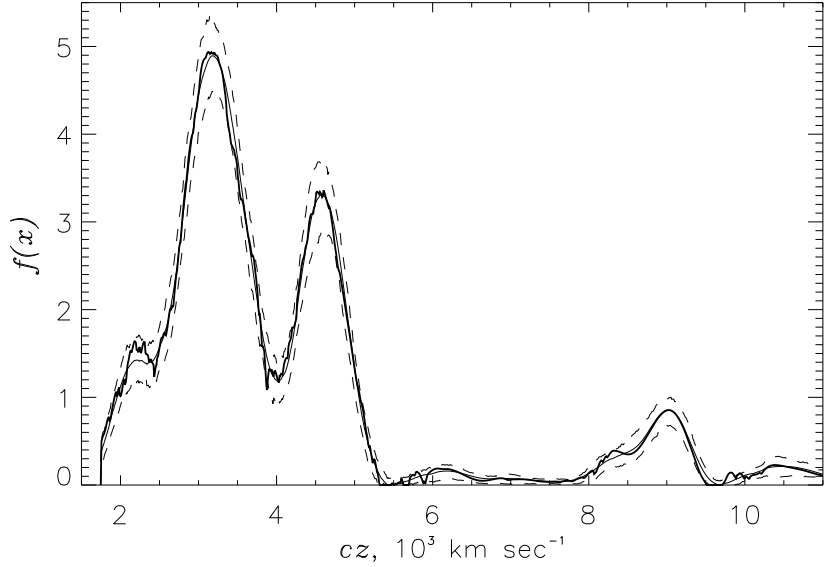}
\end{center}
\caption{The spline histogram of A3526. The meaning of lines is the same as
in Fig.~\protect\ref{sim500}}
\label{a3526}
\end{figure}

Result of the implementation of the TCSplin algorithm to the data of \cite%
{Fabricant}, consisting of 89 galaxy redshift measurements, is shown in Fig.~%
\ref{a2256}. From the figure one can obviously see that the cluster is
unrelaxed and has strongly non-Gaussian velocity PDF. Most likely it
consists of two or more merging parts that currently are undergoing a final
stage of unification.

Centaurus cluster A3526 ($z \approx 0.011$, $\alpha \approx 12^{h}48.9^{m}$, 
$\delta \approx -41^\circ18^{\prime}$, equinox 2000.0 \cite{Abell}) has been
extensively studied, as it is a nearby rich cluster of galaxies. It is
intermediate between Coma and Virgo clusters in richness and in distance and
has richness class 1 or 2 (e.g. \cite{Sandage1973}). Centaurus is irregular
in appearance, like Virgo. The cluster core has two apparent centres of
concentration, one being centred on NGC 4696 and the other being $0.5^{\circ}
$ further east (\cite{Klemola1969}, \cite{Bahcall1974}).

Extensive study of this cluster is made in \cite{Dickens1}, \cite{Dickens2}
and \cite{Dickens3}. The
research included determination of redshifts for 259 galaxies and photometry
for 329 galaxies within $13^{\circ }$ field centred on the cluster, and the
following analysis of data. The bimodal galaxy velocity distribution and
extensive substructure in both subclusters have been found. Mean
heliocentric velocities and line-of-sight dispersions of two main cluster
components, within $3^{\circ }$ of the cluster centre, are 3041 and 586 km
sec$^{-1}$ (denoted Cen30), and 4570 and 262 km sec$^{-1}$ (denoted Cen45),
respectively. The projected distributions of members of each component
overlap on the sky. Other small galaxy groups also have been found in this
study.

Recently bimodality of the cluster has been confirmed in \cite{Stein}. The
authors used it to test a non-parametric method of the PDF estimation
proposed by \cite{Fadda} and the same two main features of the cluster were
noticed.

Our result of processing the data set of \cite{Dickens1} is shown in Fig.~%
\ref{a3526}. We find the same two main structures as in the original
analysis. Clearness of these features demonstrates the quality of the
algorithm. Shape of each of these components is close to Gaussian indicating
their relatively relaxed state. Besides that the spline histogram shows
additional left ``shoulder'' of the Cen30 group at around $cz\approx 2100$
km sec$^{-1}$. This probably is one of separate groups noticed by \cite%
{Dickens3}. Possibly this as well as those features around 6200 and 8300
km~sec$^{-1}$ are not spatially real but just the redshift space caustics
artefacts.

We see that a direct implementation of the algorithm leads to a good
estimate of the PDF of clusters of galaxies. The only difference is the
dispersions of the group velocities that are overestimated due to our PDF
smoothing. One should keep that in mind and calculate the dispersion
directly from the original data if needed.

\section{Concluding remarks}

\label{beigas}

This paper has demonstrated the usefulness of the spline histogram algorithm 
in statistical studies of 1D data sets. It has all advantages over the well
known ordinary histogram approach estimating the probability density
functions. In principle the spline histograms may be expanded to higher
dimensional cases but that introduces higher effect of the sampling noise.
Unfortunately enlargement of a data set size does not necessarily guarantee
larger signal to noise ratio. More generally it is dependent on the
distribution character.


The latest version of the spline histogram algorithm TCSplin code is freely
available online from http://www.virac.lv/en/soft.html. Presently it is
written in f77 with IDL and Gnuplot visualization scripts.

\vspace{0.3cm} \noindent \textbf{Acknowledgments.} DD is grateful to the
European Physical Society EWTF that has provided a travel grant and the LOC
of Research centre of Astronomy of Academy of Athens for possibility to
participate in the Workshop. DD and KB also thank Bernard Jones for the
fruitful discussions. The work of KB was supported by grant No.01.0024.4.1
of the Latvian Council of Sciences.

%

\end{document}